\documentclass[12pt,a4paper]{article}
\usepackage[left=1in,right=1in,top=1in,bottom=1in]{geometry}
\usepackage{graphicx}
\newcommand{\beq}{\begin{eqnarray}}
\newcommand{\eeq}{\end{eqnarray}}
\newcommand{\nn}{\nonumber}
\newcommand{\pv}{{\rm pv}\,}
\newcommand{\supp}{{\rm supp}\,}
\title{$q$-Moments remove the degeneracy associated with the inversion of the $q$-Fourier transform}
\author{M. Jauregui$^1$, C. Tsallis$^{1,2}$ and E.M.F. Curado$^{1,3}$}
\date{$^1$ {\it Centro Brasileiro de Pesquisas Fisicas and National Institute of Science and Technology for Complex Systems, Rua Xavier Sigaud 150, 22290-180 Rio de Janeiro, Brazil.}\\
$^2$ {\it Santa Fe Institute, 1399 Hyde Park Road, Santa Fe, NM 87501, USA.}\\
$^3$ {\it Laboratoire APC, Universit\'e Paris Diderot, 10, rue A. Domon et L. Duquet, 75205, Paris France.}}
\begin{document}
\maketitle

\begin{abstract}It was recently proven [Hilhorst, JSTAT, P10023 (2010)] that the $q$-generalization of the Fourier transform is not invertible in the full space of probability density functions for $q>1$. It has also been recently shown that this complication disappears if we dispose of the $q$-Fourier transform not only of the function itself, but also of all of its shifts [Jauregui and Tsallis, Phys. Lett. A {\bf 375}, 2085 (2011)]. 
Here we show that another road exists for completely removing the degeneracy associated with the inversion of the $q$-Fourier transform of a given probability density function. Indeed, it is possible to determine this density if we dispose of some extra information related to its $q$-moments.
\end{abstract}

\section{Introduction}
Nonextensive statistical mechanics \cite{Tsallis1988}, a current generalization of the Boltzmann-Gibbs theory, is actively studied in diverse areas of physics and other sciences \cite{Gell-MannTsallis2004,Tsallis2009}. This theory is based on a nonadditive entropy, commonly denoted by $S_q$, that depends, in addition to the probabilities of the microstates, on a real parameter $q$, which is inherent to the system and makes $S_q$ extensive. In the limit $q\to 1$, nonextensive statistical mechanics yields the Boltzmann-Gibbs theory. This new theory has successfully described many physical and computational experiments. Such systems typically are nonergodic ones, with long-range interactions, long memory and/or other nontrivial ingredients: see for example \cite{DouglasBergaminiRenzoni2006,LiuGoree2008,PickupCywinskiPappasFaragoFouquet2009,DeVoe2009,CMS2,ALICE,Sotolongo-GrauRodriguez-PerezAntoranzSotolongo-Costa2010,AndradeSilvaMoreiraNobreCurado2010,NobreMonteiroTsallis2011}. 

The development of nonextensive statistical mechanics introduced, in addition to the generalization of some physical concepts like the Boltzmann-Gibbs-Shannon-von Neumann entropy, the generalization of some mathematical concepts. Remarkable ones are the generalizations of the classical central limit theorem and the L\'evy-Gnedenko one. These extensions are based on a generalization of the Fourier transform (FT), namely the $q$-Fourier transform ($q$-FT) \cite{UmarovTsallisSteinberg2008,UmarovTsallisGellMannSteinberg2010}. These generalized theorems respectively establish, for $q>1$, $q$-Gaussians and $(q,\alpha)$-stable distributions as attractors when the considered random variables are correlated in a special manner. 

If $1<q<3$, a $q$-Gaussian is a generalization of a Gaussian defined as a function $G_{q,\beta}:{\cal R}\to{\cal R}$ such that
\beq
\label{qG}
G_{q,\beta}(x)=\frac{\sqrt{\beta}}{C_q[1+(q-1)\beta x^2]^\frac{1}{q-1}}\equiv\frac{\sqrt{\beta}}{C_q}\exp_q(-\beta x^2)\,,
\eeq
where $\beta>0$ and $C_q$ is a normalization constant given by
\beq
\label{Cq}
C_q=\frac{\sqrt{\pi}\Gamma(\frac{3-q}{2(q-1)})}{\sqrt{q-1}\Gamma(\frac{1}{q-1})}\,.
\eeq
 A  $q$-Gaussian is not normalizable for $q \ge 3$. Its variance is finite for $q<5/3$; above this value, it diverges. When correlations can be neglected, $q\to 1$, and $G_{q,\beta}(x)\to (\beta/\pi)^{1/2}\exp(-\beta x^2)$, which is a Gaussian.

The $q$-FT of a non-negative measurable function $f$ with support $\supp f\subset{\cal R}$, denoted by $F_q[f]$, is defined, for $1 \le q <3$, as
\beq
\label{qFT}
F_q[f](\xi)=\int_{\supp f} f(x)\exp_q(i\xi x[f(x)]^{q-1})\,dx\,,
\eeq
where $\exp_q(ix)=\pv[1+(1-q)ix]^{1/(1-q)}$ for any real number $x$, being $\pv$ the notation for {\it principal value}. This is a non-linear integral transform when $q>1$. Its relevance in \cite{UmarovTsallisSteinberg2008} is that it transforms a $q$-Gaussian into another one. Hence the $q$-FT is invertible in the space of $q$-Gaussians \cite{UmarovTsallis2008}. However, it was recently proven, by means of counterexamples, that the $q$-FT is not invertible in the full space of probability density functions (pdf's) \cite{Hilhorst2010}. In connection to this problem, it is worthy mentioning that it has been found an interesting property of the $q$-FT which enables the determination of a given pdf from the knowledge of the $q$-FT of an arbitrary translation of such pdf \cite{JaureguiTsallis2011}.

Here we will discuss the counterexamples given in \cite{Hilhorst2010}, and we will show that it is possible to determine the pdf's considered in the counterexamples from the knowledge of their $q$-FT and some extra information related with their $q$-moments, defined here below.

Let $Q$ be a real number and $f$ be a pdf of some random variable $X$ such that the quantity
\beq
\nu_Q[f]=\int_{\supp f} [f(x)]^Q\,dx
\eeq
is finite. Then, we can define an {\it escort} pdf \cite{BeckSchlogl1993} for $X$, denoted by $f_Q$, as follows:
\beq
f_Q(x)=\frac{[f(x)]^Q}{\nu_Q[f]}\,.
\eeq
The moments of $f_Q$, which are called $Q$-moments of $f$, are given by
\beq
\Pi_Q^{(n)}[f]=\int_{\supp f} x^nf_Q(x)\,dx=\frac{\mu_Q^{(n)}[f]}{\nu_Q[f]}\,,
\eeq
where $\mu_Q^{(n)}[f]$ is the {\it unnormalized} $n$th $Q$-moment of $f$, defined as follows:
\beq
\mu_Q^{(n)}[f]=\int_{\supp f} x^n[f(x)]^Q\,dx\,,
\eeq
$n$ being a positive integer.

The characteristic function of $X$ is basically given by the Fourier transform of $f$, $F[f]$. It is well known that all the moments of $f$ can be obtained from the successive derivatives of the characteristic function of $X$ at the origin. It was shown that the successive derivatives of the $q$-FT of $f$ at the origin are related to specific unnormalized $Q$-moments of $f$ by the following equation \cite{TsallisPlastinoAlvarez-Estrada2009}:
\beq
\label{qFT.moments}
\left.\frac{d^nF_q[f](\xi)}{d\xi^n}\right|_{\xi=0}=i^n\left\{\prod_{j=1}^{n-1}[1+j(q-1)]\right\}\mu_{q_n}^{(n)}[f]\,,
\eeq
where $q_n=nq-(n-1)$. We can see from this relation that, if the $q$-FT of $f$ does not depend on a certain parameter that appears in $f$, then the unnormalized $n$th $q_n$-moments also do not depend on such parameter. Therefore, these unnormalized moments are unable to identify the pdf $f$ from its $q$-FT. As it will become soon clear, this difficulty does {\it not} exist for the set of $\{\nu_q \}$, which will then provide the desired identification procedure.

\section{Hilhorst's examples}
We discuss in this section two examples proposed by Hilhorst \cite{Hilhorst2010}, where the pdf depends on a certain real parameter, which disappears when we take its $q$-FT. Therefore, at the step of looking at the inverse $q$-FT, we face an infinite degeneracy. Next we illustrate, in both examples, how the degeneracy is removed through the values of the $\{\nu_q\}$. 

\subsection{First example}
\begin{figure}[htp]
\centering
\includegraphics[width=0.5\textwidth,keepaspectratio]{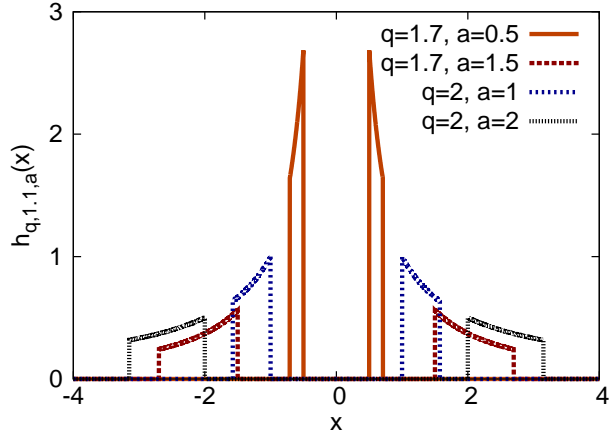}
\caption{Representation of $h_{q,\lambda,a}$ for $\lambda=1.1$ and different values of $q$ and $a$.}
\label{f.h}
\end{figure}

Let us consider the function $h_{q,\lambda,a}:{\cal R}\to{\cal R}$ such that \cite{Hilhorst2010}
\beq
h_{q,\lambda,a}(x)=\left(\frac{\lambda}{|x|}\right)^\frac{1}{q-1}
\eeq
if $a<|x|<b$, where $q>1$, and $(a,b,\lambda)$ are positive real numbers; otherwise $h_{q,\lambda,a}(x)=0$ (see Fig. \ref{f.h}). We can impose the following normalization condition for this function:
\beq
\label{h.norm}
\int_{-\infty}^{+\infty} h_{q,\lambda,a}(x)\,dx=1\,.
\eeq
From this, it follows that one parameter among $q,\lambda,a,b$ depends on the other ones. Choosing $b$ as the dependent parameter, we get
\beq
\label{h.b}
b=
\left\{\begin{array}{ll}
[\frac{q-2}{2(q-1)}\lambda^\frac{1}{1-q}+a^\frac{q-2}{q-1}]^\frac{q-1}{q-2}& \mbox{if $q\ne2$}\\
ae^\frac{1}{2\lambda}& \mbox{if $q=2$.}
\end{array}\right.
\eeq

Given $Q$ such that $1\le Q<3$, the $Q$-FT of $h_{q,\lambda,a}$ can be easily reduced to the following expression:
\beq
\label{hqFT}
F_Q[h_{q,\lambda,a}](\xi)=2\int_a^b\left(\frac{\lambda}{x}\right)^\frac{1}{q-1}\cos_Q\left(\xi x\left(\frac{\lambda}{x}\right)^\frac{Q-1}{q-1}\right)\,dx\,,
\eeq
where $\cos_q$ is the $q$-generalization of the trigonometric function $\cos$ which is defined by \cite{Borges1998}
\beq
\cos_q(x)=\Re(\exp_q(ix))=\frac{\cos(\frac{1}{q-1}\arctan((q-1)x))}{[1+(q-1)^2x^2]^\frac{1}{2(q-1)}}\,.
\eeq

\begin{figure}[htp]
\centering
\includegraphics[width=0.5\textwidth,keepaspectratio]{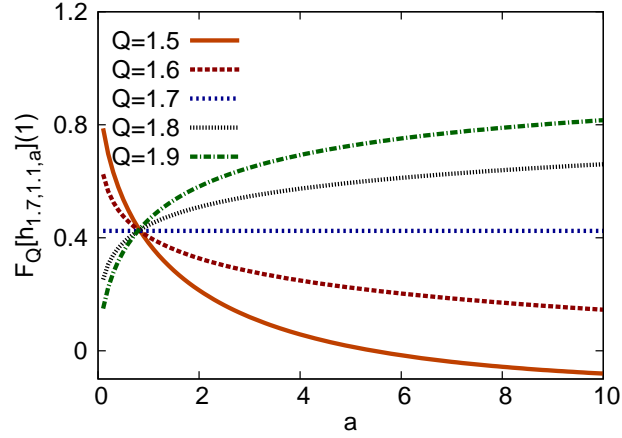}
\caption{The dependence on $a$ of $F_Q[h_{1.7,1.1,a}](1)$ for different values of $Q$.}
\label{f.hqFT}
\end{figure}

It is easy to notice from (\ref{hqFT}) that the $Q$-FT of $h_{q,\lambda,a}$ depends on $a$ if $Q \ne q$. However, it does not depend on $a$ when $Q=q$ (see Fig. \ref{f.hqFT}), when it is given by $F_q[h_{q,\lambda,a}](\xi)=\cos_q(\xi\lambda)$.

Consequently, there exist infinite functions $h_{q,\lambda,a}$ with the same $q$ and $\lambda$ but different $a$, which have the same $q$-FT. Therefore, it is not possible to determine $h_{q,\lambda,a}$ just from the knowledge of its $q$-FT. However, it may be possible to obtain $h_{q,\lambda,a}$ from its $q$-FT {\it and} some extra information. For example, we would be able to determine $h_{q,\lambda,a}$ if we knew the $q$-FT of an arbitrary translation of $h_{q,\lambda,a}$ \cite{JaureguiTsallis2011}. Here we will give another approach to this problem.

As $h_{q,\lambda,a}$ is a non-negative function, which obeys the normalization condition (\ref{h.norm}), it can be interpreted as a pdf of some random variable. Moreover, for any real number $Q$, we have that
\beq
\label{h.nuQ}
\nu_Q[h_{q,\lambda,a}]=
\left\{\begin{array}{ll}
\frac{2(q-1)}{q-1-Q}\lambda^\frac{Q}{q-1}(b^\frac{q-1-Q}{q-1}-a^\frac{q-1-Q}{q-1})& \mbox{if $Q\ne q-1$}\\
2\lambda\ln(\frac{b}{a})& \mbox{if $Q=q-1$}
\end{array}
\right.
\eeq
is finite. Being $n$ an even positive integer, we have also that the unnormalized $n$th $Q$-moment of $h_{q,\lambda,a}$ is given by
\beq
\mu_Q^{(n)}[h_{q,\lambda,a}]=
\left\{\begin{array}{ll}
\frac{2(q-1)}{(n+1)(q-1)-Q}\lambda^\frac{Q}{q-1}[b^\frac{(n+1)(q-1)-Q}{q-1}-a^\frac{(n+1)(q-1)-Q}{q-1}]& \mbox{if $Q\ne (n+1)(q-1)$}\\
2\lambda^{n+1}\ln(\frac{b}{a})& \mbox{if $Q=(n+1)(q-1)$.}
\end{array}
\right.
\eeq
Then, finally, the $n$th $Q$-moment of $h_{q,\lambda,a}$ is given by
\beq
\Pi_Q^{(n)}[h_{q,\lambda,a}]=
\left\{\begin{array}{ll}
(b^n-a^n)[n\ln(\frac{b}{a})]^{-1}& \mbox{if $Q=q-1$}\\
\frac{na^nb^n}{b^n-a^n}\ln(\frac{b}{a})& \mbox{if $Q=(n+1)(q-1)$}\\
\frac{(q-1-Q)}{(n+1)(q-1)-Q}\left[\frac{b^\frac{(n+1)(q-1)-Q}{q-1}-a^\frac{(n+1)(q-1)-Q}{q-1}}{b^\frac{q-1-Q}{q-1}-a^\frac{q-1-Q}{q-1}}\right]&\mbox{otherwise.}
\end{array}
\right.
\eeq
It is clear that $\mu_Q^{(m)}[h_{q,\lambda,a}]=0$ and $\Pi_Q^{(m)}[h_{q,\lambda,a}]=0$ for any odd positive integer $m$, since $h_{q,\lambda,a}(x)$ is an even function.

As the $q$-FT of $h_{q,\lambda,a}$ does not depend on $a$, then, according to (\ref{qFT.moments}), the $n$th $q_n$-moment of $h_{q,\lambda,a}$ does not depend on $a$ either, where $q_n=nq-(n-1)$. In fact, if $q\ne 2$, we have that
\beq
\mu_{q_n}^{(n)}[h_{q,\lambda,a}]=\frac{2(q-1)}{q-2}\lambda^\frac{nq-(n-1)}{q-1}(b^\frac{q-2}{q-1}-a^\frac{q-2}{q-1})\,.
\eeq
Then, using (\ref{h.b}), we obtain that $\mu_{q_n}^{(n)}[h_{q,\lambda,a}]=\lambda^n$. If $q=2$, we have that $\mu_{n+1}^{(n)}[h_{q,\lambda,a}]=2\lambda^{n+1}\ln(b/a)$, and, using (\ref{h.b}), we obtain that $\mu_{n+1}^{(n)}[h_{q,\lambda,a}]=\lambda^n$.

\begin{figure}[htp]
\centering
\begin{tabular}{cc}
\includegraphics[width=0.45\textwidth,keepaspectratio]{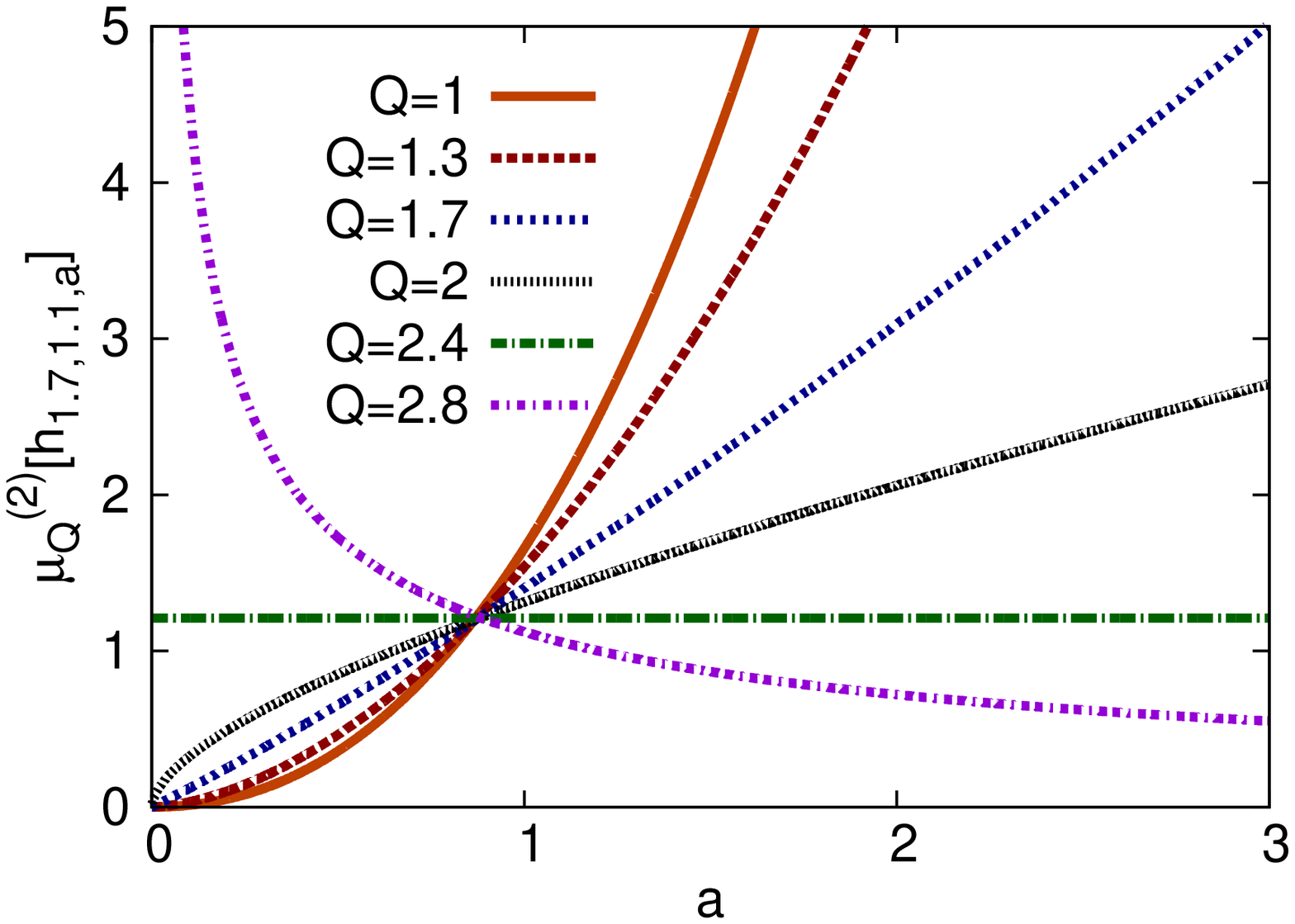} & \includegraphics[width=0.45\textwidth,keepaspectratio]{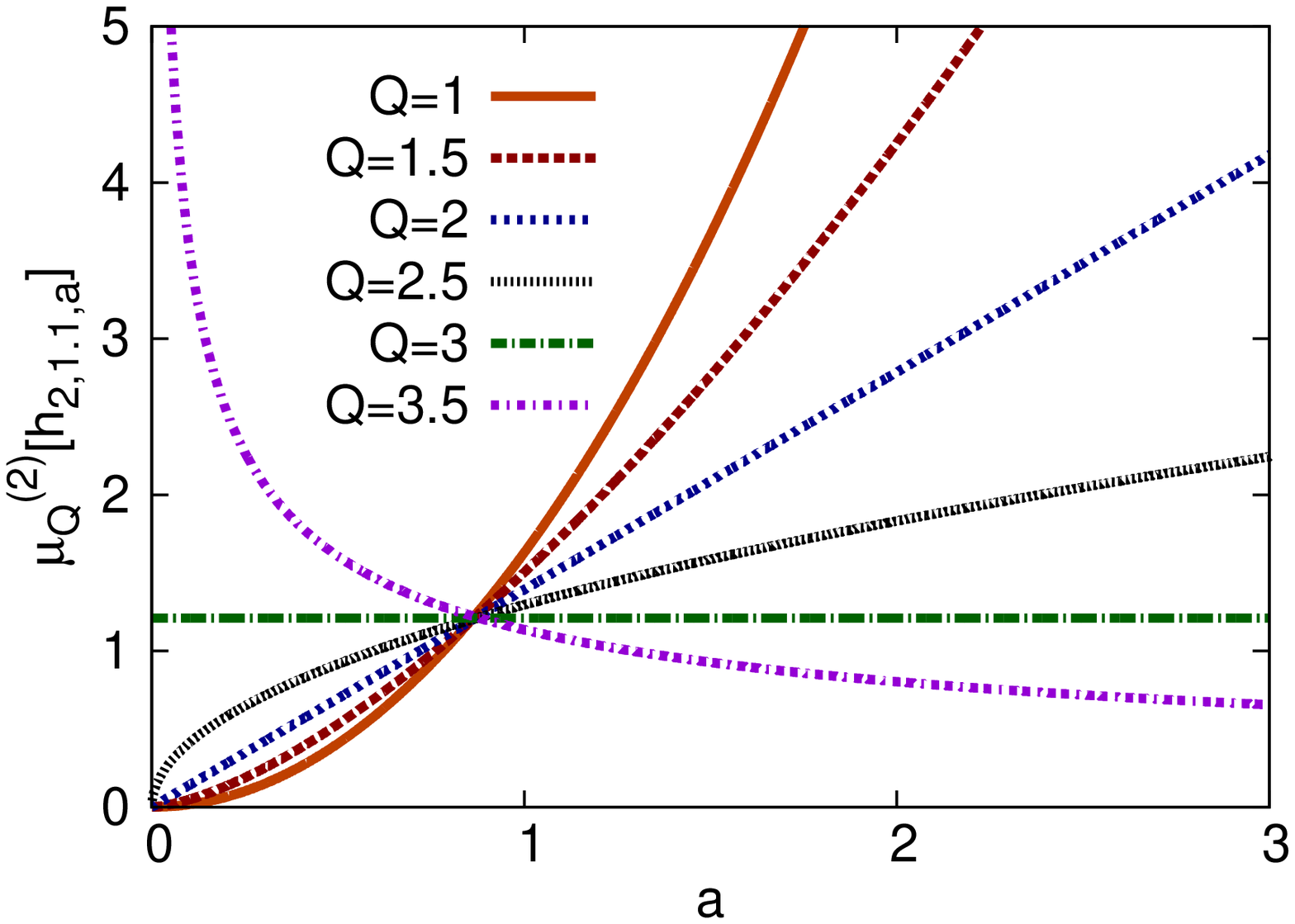}\\
(a) & (b)
\end{tabular}
\caption{The dependence on $a$ of the quantities (a) $\mu_Q^{(2)}[h_{1.7,1.1,a}]$ and (b) $\mu_Q^{(2)}[h_{2,1.1,a}]$ for different values of $Q$.}
\label{f.hmuQ}
\end{figure}

\begin{figure}[htp]
\centering
\begin{tabular}{cc}
\includegraphics[width=0.45\textwidth,keepaspectratio]{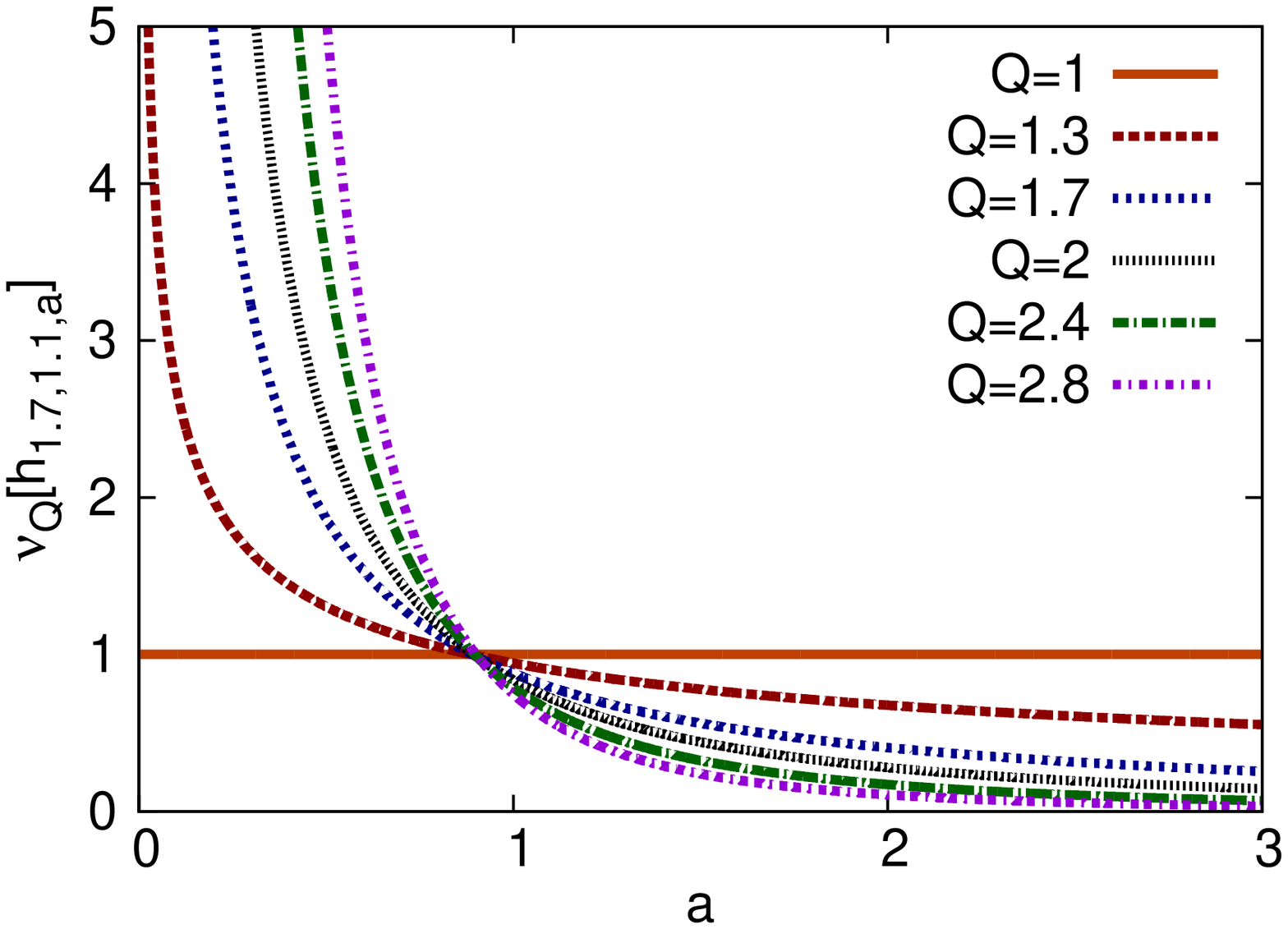} & \includegraphics[width=0.45\textwidth,keepaspectratio]{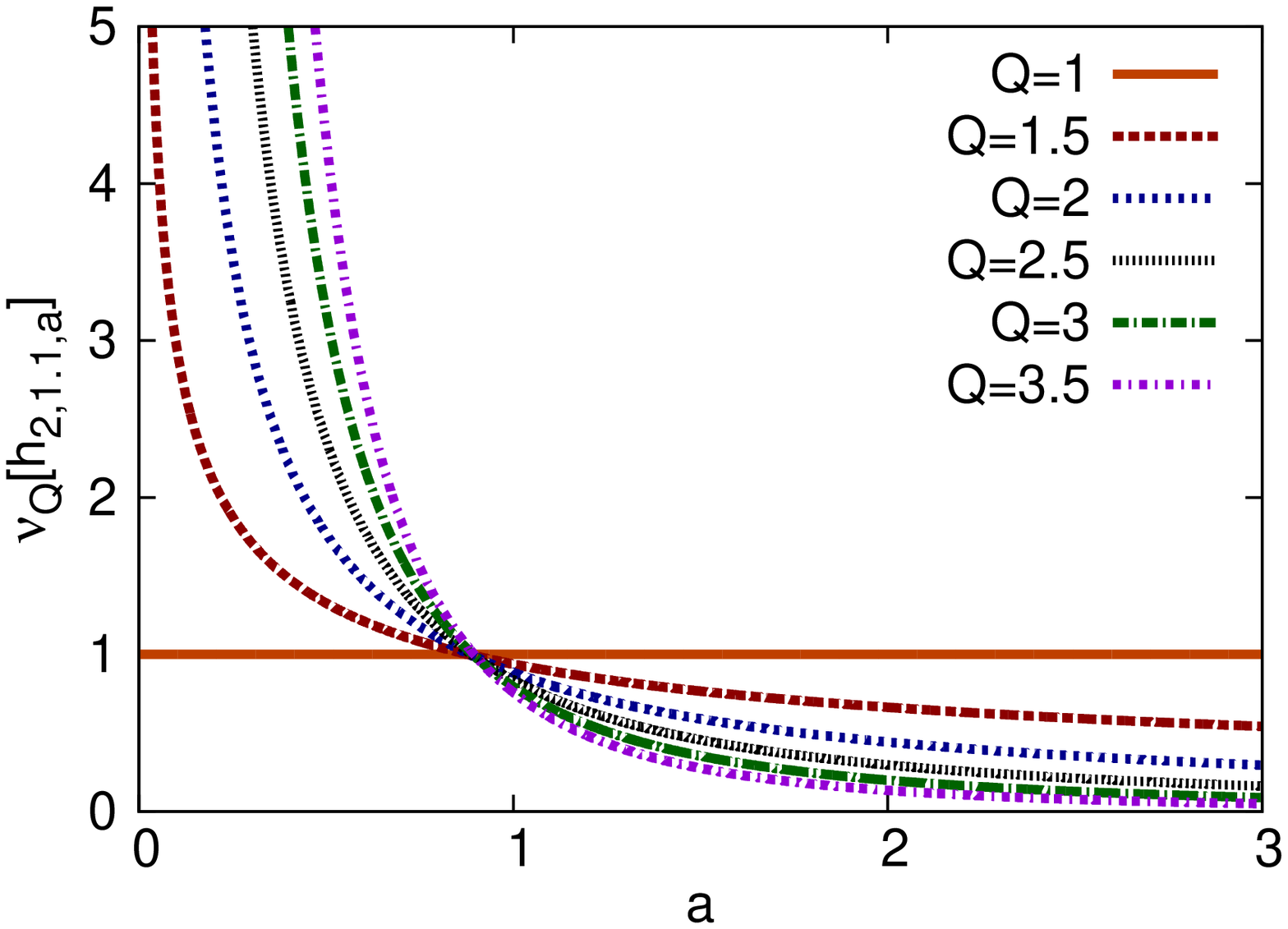}\\
(a) & (b)
\end{tabular}
\caption{The dependence on $a$ of the quantities (a) $\nu_Q[h_{1.7,1.1,a}]$ and (b) $\nu_Q[h_{2,1.1,a}]$ for different values of $Q$.}
\label{f.hnuQ}
\end{figure}

\begin{figure}[htp]
\centering
\begin{tabular}{cc}
\includegraphics[width=0.45\textwidth,keepaspectratio]{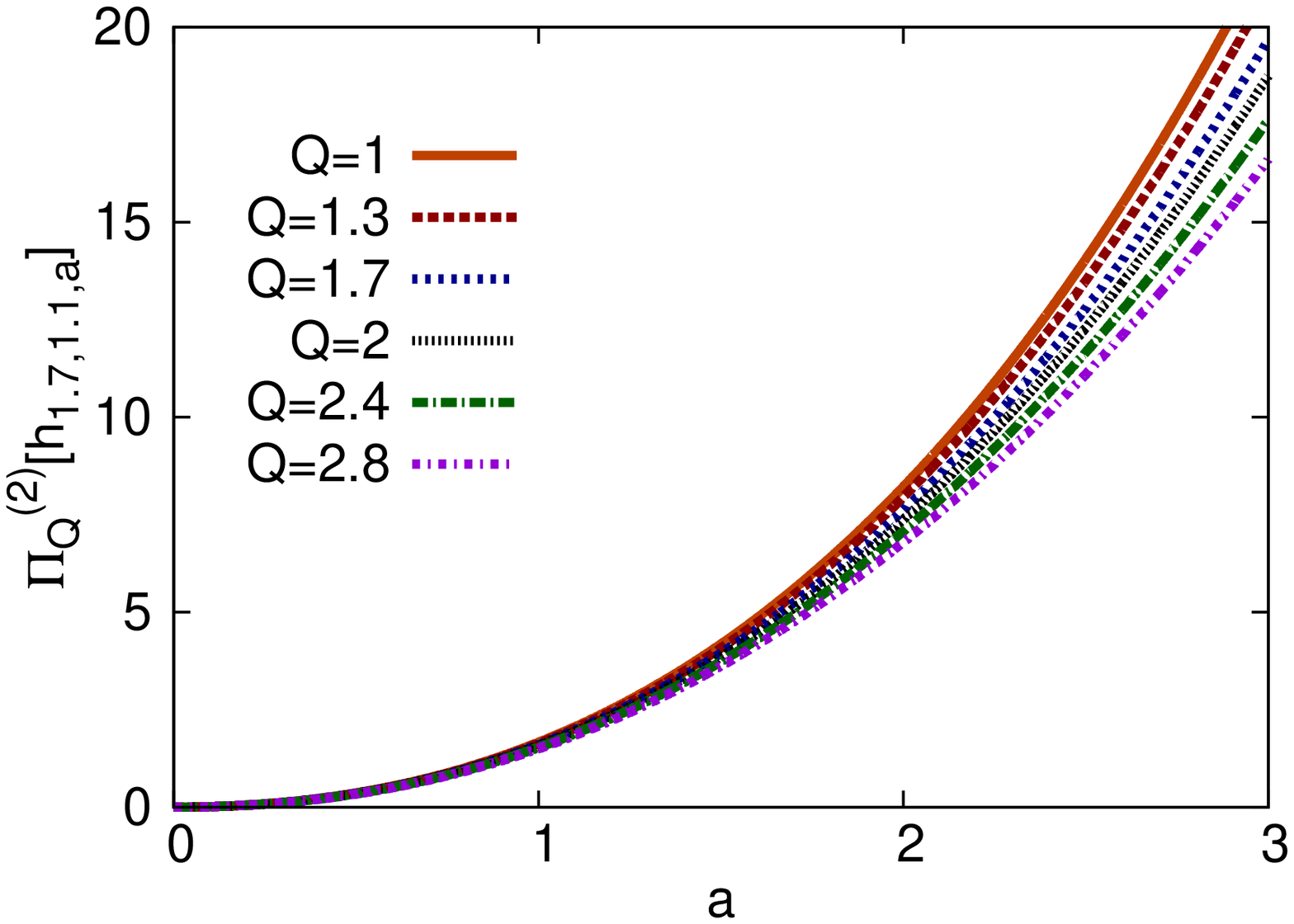} & \includegraphics[width=0.45\textwidth,keepaspectratio]{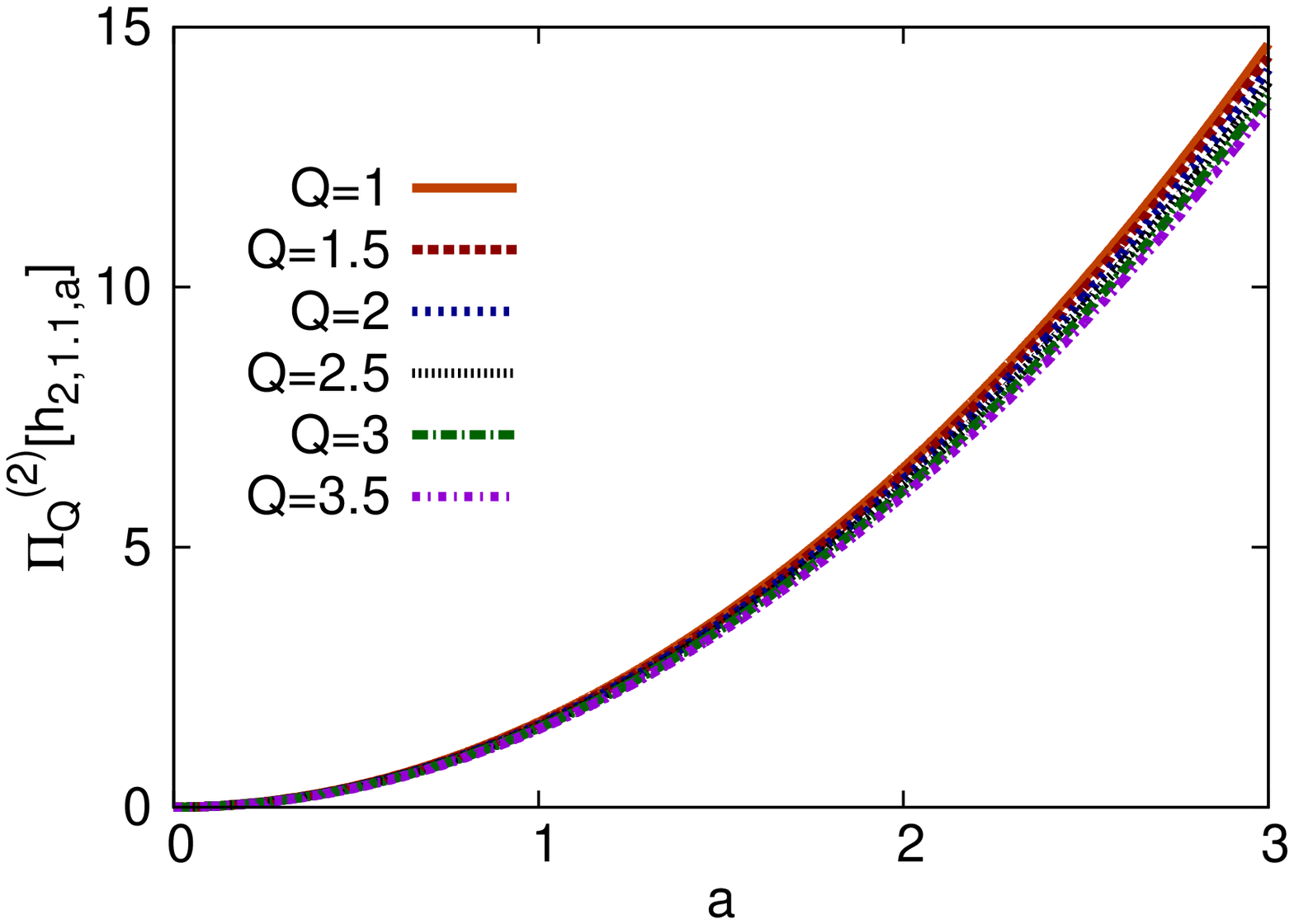}\\
(a) & (b)
\end{tabular}
\caption{The dependence on $a$ of the quantities (a) $\Pi_Q^{(2)}[h_{1.7,1.1,a}]$ and (b) $\Pi_Q^{(2)}[h_{2,1.1,a}]$ for different values of $Q$.}
\label{f.hPiQ}
\end{figure}

While the unnormalized $Q$-moments of $h_{q,\lambda,a}$ may not depend on $a$ (see Fig. \ref{f.hmuQ}), we can straightforwardly verify from (\ref{h.nuQ}) that the quantity $\nu_Q[h_{q,\lambda,a}]$ depends monotonically on $a$ for any $Q\ne 1$ (see Fig. \ref{f.hnuQ}). The same is true for the normalized $Q$-moments (see Fig. \ref{f.hPiQ}). Hence, the knowledge of the $q$-FT of $h_{q,\lambda,a}$ and the value of some $\nu_Q[h_{q,\lambda,a}]$ with $Q\ne 1$ (extra information) is sufficient to determine the pdf $h_{q,\lambda,a}$. We should notice that $\nu_1[h_{q,\lambda,a}]=1$ (it does not depend on $a$), then the extra information in this case is trivial.

\subsection{Second example}
\begin{figure}[htp]
\centering
\includegraphics[width=0.5\textwidth,keepaspectratio]{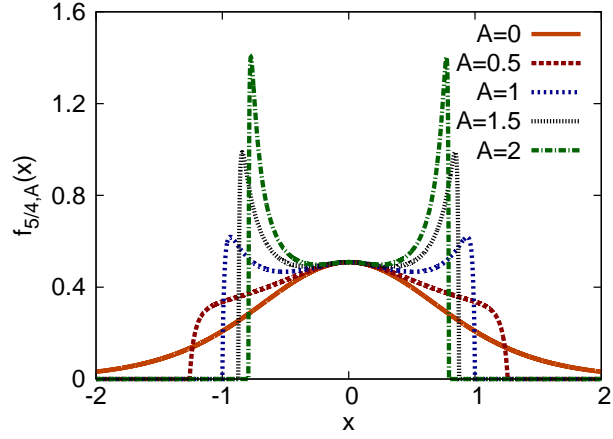}
\caption{Representation of $f_{5/4,A}$ for different values of $A$.}
\label{f.f}
\end{figure}

Let us consider now the function $f_{q,A}:{\cal R}\to{\cal R}$ such that \cite{Hilhorst2010}
\beq
f_{q,A}(x)=\frac{(1-A|x|^\frac{2-q}{q-1})^\frac{1}{q-2}}{C_q[1+(q-1)x^2(1-A|x|^\frac{2-q}{q-1})^\frac{2(q-1)}{q-2}]^\frac{1}{q-1}}
\eeq
if $|x|^{(q-2)/(q-1)}>A$, where $1<q<2$, $A\ge 0$, and $C_q$ is the normalization constant of a $q$-Gaussian given by (\ref{Cq}); otherwise $f_{q,A}(x)=0$ (see Fig. \ref{f.f}). We can easily notice that $f_{q,0}(x)=G_{q,1}(x)$, where $G_{q,\beta}(x)$ is defined in (\ref{qG}).

\begin{figure}[htp]
\centering
\includegraphics[width=0.5\textwidth,keepaspectratio]{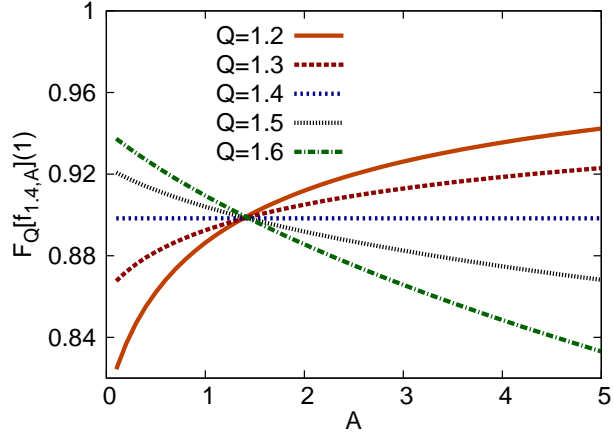}
\caption{The dependence on $A$ of $F_Q[f_{1.4,A}](1)$ for different values of $Q$.}
\label{f.fqFT}
\end{figure}

Let $1<Q<3$ and $A>0$. The $Q$-FT of $f_{q,A}$ is given by (see Fig. \ref{f.fqFT})
\beq
F_Q[f_{q,A}](\xi)=\int_{-A^\frac{q-1}{q-2}}^{A^\frac{q-1}{q-2}} f_{q,A}(x)\exp_Q(i\xi x[f_{q,A}(x)]^{Q-1})\,dx\,.
\eeq
\noindent In order to compute this integral in the particular case $Q=q$, we should notice first that
\beq
\exp_q(i\xi x[f_{q,A}(x)]^{q-1})&=&\exp_q\left(\frac{i\xi x(1-A|x|^\frac{2-q}{q-1})^\frac{q-1}{q-2}}{C_q^{q-1}[1+(q-1)x^2(1-A|x|^\frac{2-q}{q-1})^\frac{2(q-1)}{q-2}]}\right)\nn\\
&=&\frac{[1+(q-1)x^2(1-A|x|^\frac{2-q}{q-1})^\frac{2(q-1)}{q-2}]^\frac{1}{q-1}}{\pv\left\{1-(q-1)\left[\frac{-x^2}{(1-A|x|^\frac{2-q}{q-1})^\frac{2(q-1)}{2-q}}+\frac{iC_q^{1-q}\xi x}{(1-A|x|^\frac{2-q}{q-1})^\frac{q-1}{2-q}}\right]\right\}^\frac{1}{q-1}}\nn\\
&=&[1+(q-1)x^2(1-A|x|^\frac{2-q}{q-1})^\frac{2(q-1)}{q-2}]^\frac{1}{q-1}\nn\\
&&\times\exp_q\left(\frac{-x^2}{(1-A|x|^\frac{2-q}{q-1})^\frac{2(q-1)}{2-q}}+\frac{iC_q^{1-q}\xi x}{(1-A|x|^\frac{2-q}{q-1})^\frac{q-1}{2-q}}\right)\,.
\eeq
Then
\beq
F_q[f_{q,A}](\xi)&=&\frac{1}{C_q}\int_{-A^\frac{q-1}{q-2}}^{A^\frac{q-1}{q-2}} \frac{\exp_q\left(\frac{-x^2}{(1-A|x|^\frac{2-q}{q-1})^\frac{2(q-1)}{2-q}}+\frac{iC_q^{1-q}\xi x}{(1-A|x|^\frac{2-q}{q-1})^\frac{q-1}{2-q}}\right)}{(1-A|x|^\frac{2-q}{q-1})^\frac{1}{2-q}}\,dx\nn\\
&=&\frac{1}{C_q}\int_{-A^\frac{q-1}{q-2}}^{A^\frac{q-1}{q-2}} \frac{\exp_q\left(-\left[\frac{x}{(1-A|x|^\frac{2-q}{q-1})^\frac{q-1}{2-q}}-\frac{iC_q^{1-q}\xi}{2}\right]^2-\frac{C_q^{2(1-q)}\xi^2}{4}\right)}{(1-A|x|^\frac{2-q}{q-1})^\frac{1}{2-q}}\,dx\,.
\eeq
Finally, using the change of variables
\beq
y=\frac{x}{(1-A|x|^\frac{2-q}{q-1})^\frac{q-1}{2-q}}-\frac{iC_q^{1-q}\xi}{2}\,,
\eeq
we obtain that
\beq
\label{fqFT}
F_q[f_{q,A}](\xi)&=&\frac{1}{C_q}\int_{-\infty-\frac{iC_q^{1-q}\xi}{2}}^{+\infty-\frac{iC_q^{1-q}\xi}{2}} \exp_q\left(-y^2-\frac{C_q^{2(1-q)}\xi^2}{4}\right)\,dy\,
\eeq
which does {\it not} depend on $A$. Moreover, the RHS of (\ref{fqFT}) is equal to the $q$-FT of the $q$-Gaussian $G_{q,1}$ (see details in \cite{UmarovTsallisSteinberg2008}), which, naturally, does not depend on $A$. Then, the knowledge of only the $q$-FT of $f_{q,A}$ would not be sufficient information to determine $f_{q,A}$. Hence, as in the first example, extra information is needed.

\begin{figure}[t]
\centering
\includegraphics[width=0.5\textwidth,keepaspectratio]{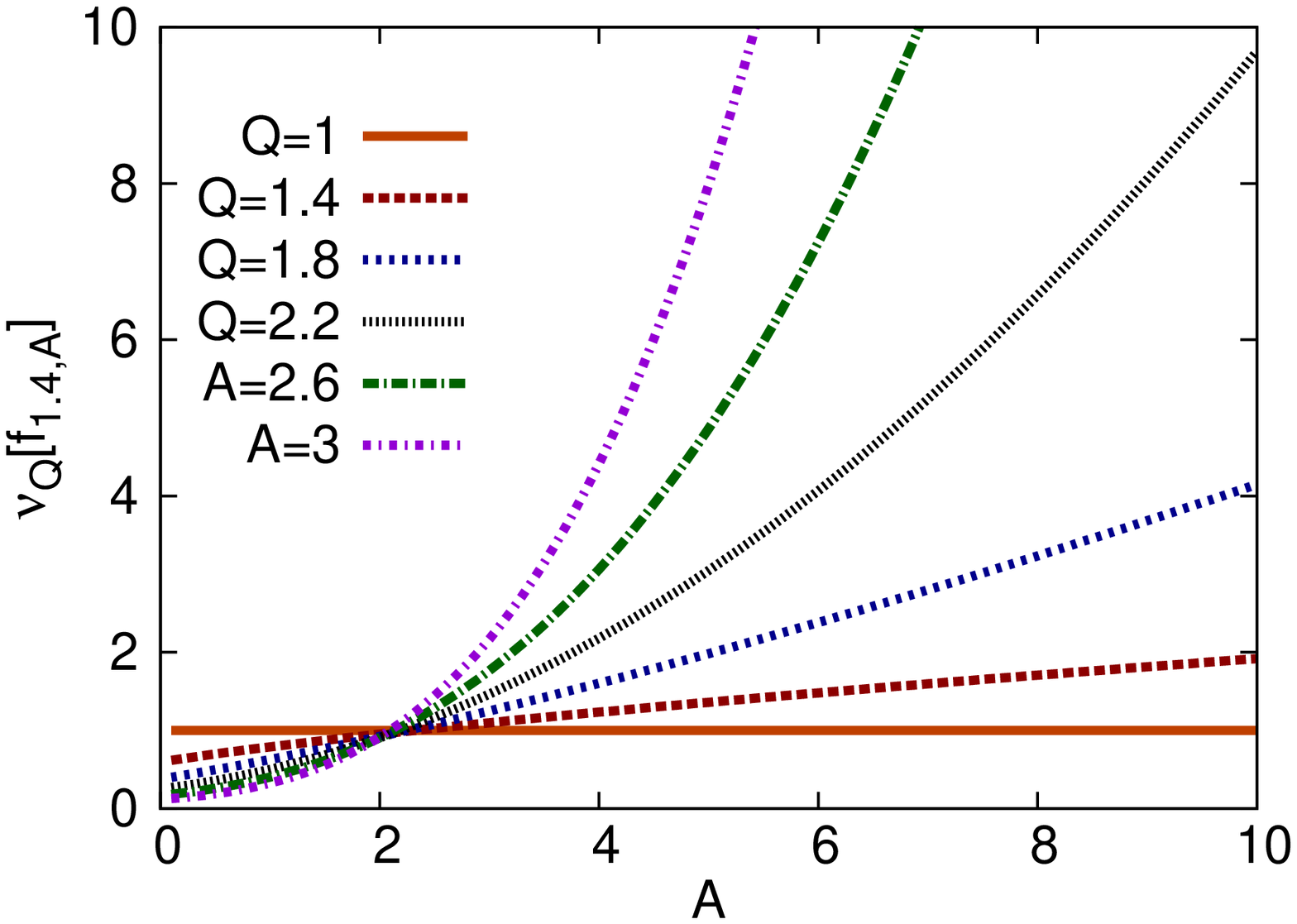}
\caption{The dependence on $A$ of the quantity $\nu_Q[f_{1.4,A}]$ for different values of $Q$.}
\label{f.fnuQ}
\end{figure}

\begin{figure}[htp]
\centering
\includegraphics[width=0.5\textwidth,keepaspectratio]{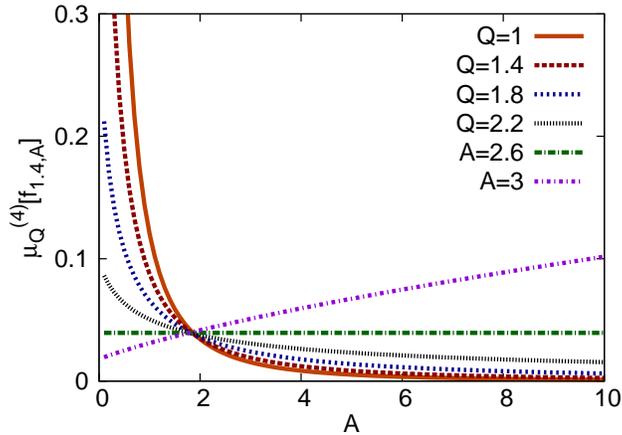}
\caption{The dependence on $A$ of the unnormalized fourth $Q$-moment of $f_{1.4,A}$ for different values of $Q$.}
\label{f.fmuQ}
\end{figure}

Let $Q$ be a real number. Considering $f_{q,A}$ as a pdf of some random variable, we have that
\beq
\nu_Q[f_{q,A}]&=&\int_{-A^\frac{q-1}{q-2}}^{A^\frac{q-1}{q-2}}\frac{(1-A|x|^\frac{2-q}{q-1})^\frac{Q}{q-2}}{C_q^Q[1+(q-1)x^2(1-A|x|^\frac{2-q}{q-1})^\frac{2(q-1)}{q-2}]^\frac{Q}{q-1}}\,dx\nn\\
&=&\frac{1}{C_q^Q}\int_{-A^\frac{q-1}{q-2}}^{A^\frac{q-1}{q-2}}\frac{\left[\exp_q\left(\frac{-x^2}{(1-A|x|^\frac{2-q}{q-1})^\frac{2(q-1)}{2-q}}\right)\right]^Q}{(1-A|x|^\frac{2-q}{q-1})^\frac{Q}{2-q}}\,dx\,,
\eeq
which is finite and depends on $A$ when $Q\ne 1$ (see Fig. \ref{f.fnuQ}).
\noindent The unnormalized $n$th $Q$-moment of $f_{q,A}$ for any positive integer $n$ is given by
\beq
\mu_Q^{(n)}[f_{q,A}]=\frac{1}{C_q^Q}\int_{-A^\frac{q-1}{q-2}}^{A^\frac{q-1}{q-2}}\frac{x^n\left[\exp_q\left(\frac{-x^2}{(1-A|x|^\frac{2-q}{q-1})^\frac{2(q-1)}{2-q}}\right)\right]^Q}{(1-A|x|^\frac{2-q}{q-1})^\frac{Q}{2-q}}\,dx\,,
\eeq
which depends on $A$ except when $Q=q_n=nq-(n-1)$ (see Fig. \ref{f.fmuQ}). In this case, using the change of variables
\beq
y=\frac{x}{(1-A|x|^\frac{2-q}{q-1})^\frac{q-1}{2-q}},
\eeq
we obtain that
\beq
\mu_{q_n}^{(n)}[f_{q,A}]=\int_{-\infty}^{+\infty} y^n\left[\frac{1}{C_q}\exp_q(-y^2)\right]^{nq-(n-1)}\,dy\,,
\eeq
which is equal to the unnormalized $n$th $q_n$-moment of the $q$-Gaussian $G_{q,1}$. Therefore, we see that, like in the first example, the knowledge of any  $\nu_Q[f_{q,A}]$ with $Q \ne 1$ enables the determination of the pdf $f_{q,A}$ from its $q$-FT.
\section{Conclusions}
Both functions $h_{q,\lambda,a}$ and $f_{q,A}$ show that the $q$-FT is not invertible in the full space of pdf's, since their $q$-FT's do not depend on $a$ and $A$ respectively. However, if $Q\ne q$, this problem would not occur for the $Q$-FT of both functions (see Figs. \ref{f.hqFT} and \ref{f.fqFT}). In other words, the $Q$-FT of both functions with $Q \ne q$ would in principle be invertible. Furthermore, in the case $Q=q$, Figs. \ref{f.hnuQ} and \ref{f.fnuQ} show that the quantities $\nu_Q[h_{q,\lambda,a}]$ and $\nu_Q[f_{q,A}]$ depend monotonically on $a$ and $A$ respectively, which removes the degeneracy. Therefore, the knowledge of the $q$-FT of both functions and a single value of $\nu_Q[h_{q,\lambda,a}]$ and $\nu_Q[f_{q,A}]$ is sufficient to determine the functions $h_{q,\lambda,a}$ and $f_{q,A}$.

If we were in the case that a pdf $f$ depends on two or more parameters and its $q$-FT does not depend on more than one of such parameters, we would expect this method of identification of the inverse $q$-FT to work as fine as in the case of the functions considered in this paper. However, it might be possible that more than one value of $\nu_Q$ is needed.
\subsection*{Acknowledgments}
We acknowledge very fruitful remarks from H.J. Hilhorst, F.D. Nobre and S. Umarov. Partial financial support from CNPq and Faperj (Brazilian Agencies) is acknowledged as well.


\begin{thebibliography}{10}

\bibitem{Tsallis1988} C. Tsallis, J. Stat. Phys. {\bf 52}, 479 (1988).
\bibitem{Gell-MannTsallis2004} M. Gell-Mann, C. Tsallis (Eds.), {\it Nonextensive Entropy -- Interdisciplinary Applications}, Oxford University Press, New York (2004).
\bibitem{Tsallis2009} C. Tsallis, {\it Introduction to Nonextensive Statistical Mechanics -- Approaching a
Complex World}, Springer, New York (2009).
\bibitem{DouglasBergaminiRenzoni2006}P. Douglas, S. Bergamini and F. Renzoni, Phys. Rev. Lett. {\bf 96}, 110601 (2006).
\bibitem{LiuGoree2008}B. Liu and J. Goree, Phys. Rev. Lett. {\bf 100}, 055003 (2008).
\bibitem{PickupCywinskiPappasFaragoFouquet2009}R.M. Pickup, R. Cywinski, C. Pappas, B. Farago and P. Fouquet, Phys. Rev. Lett. {\bf 102}, 097202 (2009).
\bibitem{DeVoe2009}R.G. DeVoe, Phys. Rev. Lett. {\bf 102}, 063001 (2009).
\bibitem{CMS2}CMS Collaboration, Phys. Rev. Lett. {\bf 105}, 022002 (2010).
\bibitem{ALICE}ALICE Collaboration, 
Eur. Phys. J. C {\bf 71}, 1594  (2011).
\bibitem{Sotolongo-GrauRodriguez-PerezAntoranzSotolongo-Costa2010}O. Sotolongo-Grau, D. Rodriguez-Perez, J.C. Antoranz and O. Sotolongo-Costa, Phys. Rev. Lett. {\bf 105}, 158105 (2010).
\bibitem{AndradeSilvaMoreiraNobreCurado2010}J. S. Andrade Jr., G.F.T. da Silva, A.A. Moreira, F.D. Nobre and E.M.F. Curado, Phys. Rev. Lett. {\bf 105}, 260601 (2010). 
\bibitem{NobreMonteiroTsallis2011}F.D. Nobre, M.A. Rego-Monteiro and C. Tsallis, Phys. Rev. Lett. {\bf 106}, 140601 (2011).
\bibitem{UmarovTsallisSteinberg2008} S. Umarov, C. Tsallis and S. Steinberg, Milan J. Math. {\bf 76}, 307 (2008).
\bibitem{UmarovTsallisGellMannSteinberg2010}S. Umarov, C. Tsallis, M. Gell-Mann and S. Steinberg,  
J. Math. Phys. {\bf 51}, 033502 (2010).  
\bibitem{UmarovTsallis2008} S. Umarov and C. Tsallis, Phys. Lett. A {\bf 372}, 4874 (2008).
\bibitem{Hilhorst2010} H.J. Hilhorst, J. Stat. Mech. P10023 (2010).
\bibitem{JaureguiTsallis2011} M. Jauregui and C. Tsallis, Phys. Lett. A {\bf 375}, 2085 (2011).
\bibitem{BeckSchlogl1993}C. Beck and F. Schlogl, {\it Thermodynamics of Chaotic Systems} (Cambridge University Press, Cambridge, 1993).
\bibitem{TsallisPlastinoAlvarez-Estrada2009} C. Tsallis, A.R. Plastino and R.F. Alvarez-Estrada, J. Math. Phys. {\bf 50}, 043303 (2009).
\bibitem{Borges1998} E.P. Borges, J. Phys. A {\bf 31}, 5281 (1998).
\end{thebibliography}
\end{document}